\documentclass[letterpaper, 10pt, conference]{ieeeconf}      

\IEEEoverridecommandlockouts                              
\usepackage[utf8]{inputenc}
\usepackage[english]{babel}

\usepackage{amsmath} 
\usepackage{amssymb}  
\usepackage{graphicx}
\graphicspath{{./figure/}}
\usepackage{subcaption}
\usepackage{color}
\usepackage{enumerate}
\usepackage{algorithm}
\usepackage{algpseudocode}
\usepackage{float}
\usepackage{stfloats}

\usepackage{pgfplots}
\pgfplotsset{width=10cm, compat=1.9, legend style={font=\footnotesize}}
\tikzset{%
	dot/.style={circle, fill=black, minimum size=4pt, inner sep=0pt, outer sep=-1pt},
	hdot/.style={circle, fill=white, minimum size=4pt, inner sep=0pt, outer sep=-1pt},
}
\newlength\fheight 
\newlength\fwidth 
\usepackage{circuitikz}
\usepgflibrary{arrows}

\usepackage{tikz}
\usetikzlibrary{calc, backgrounds}
\usetikzlibrary{shapes.geometric, arrows}

\definecolor{proc}{RGB}{16,207,155}
\tikzstyle{process}  = [rectangle, rounded corners, minimum width=2cm, minimum height=1cm,text centered, text width=2cm, draw=black, fill=proc!50]

\definecolor{dec}{RGB}{11,208,217}
\tikzstyle{decision} = [rectangle, minimum width=2cm, minimum height=1cm, text centered, text width=2.5cm, draw=black, fill=dec!50]

\tikzstyle{scenario} = [rectangle, rounded corners, minimum width=2cm, minimum height=0.5cm, text centered, text width=3cm, draw=black, fill=pink!20]

\tikzstyle{arrow}= [thick, ->, >=stealth]

\tikzstyle{system} = [rectangle, rounded corners, minimum width=4cm, minimum height=4cm, text centered, text width=4cm, draw=black, fill=red!40,fill opacity = 0.2]

\tikzstyle{controller} = [rectangle,rounded corners,minimum width=1cm,minimum height = 1cm,text centered,text width=1cm,draw=black,fill=green!50,fill opacity = 0.2]

\tikzstyle{detector} = [rectangle,rounded corners,minimum width=1cm,minimum height = 1cm,text centered,text width=1cm,draw=black,fill=yellow!50,fill opacity = 0.2]

\newcommand{\matrices}[1]{\begin{bmatrix} #1\end{bmatrix}}

\newtheorem{theorem}{Theorem}
\newtheorem{lemma}{Lemma}
\newtheorem{proposition}{Proposition}

\newtheorem{definition}{Definition}
\newtheorem{assumption}{Assumption}
\newtheorem{remark}{Remark}
\newtheorem{problem}{Problem}

\newcommand{\commentout}[1]{}

\newcommand{\PP}{\mathcal{P}}
\newcommand{\CC}{\mathcal{C}}
\newcommand{\WW}{\mathcal{W}}
\newcommand{\QQ}{\mathcal{Q}}
\newcommand{\HH}{\mathcal{H}}
\newcommand{\GG}{\mathcal{G}}

\allowdisplaybreaks

\definecolor{new}{RGB}{0,0,0}

\definecolor{fb} {RGB}{34,139,34}
\definecolor{cut}{RGB}{255,0,0}

\definecolor{ajg}{RGB}{0,0,0}

\title{\LARGE \bf
	Design of multiplicative watermarking against covert attacks
}

\author{ Alexander J. Gallo, Sribalaji C. Anand, Andr\'{e} M. H. Teixeira, Riccardo M. G. Ferrari
	\thanks{
		This work has been partially supported by the Research Council of Norway through the project AIMWind, grant id 312486.
		This work is supported by the Swedish Research Council under the grant 2018-04396 and by the Swedish Foundation for Strategic Research.}
	\thanks{S. C. Anand is with the Department of Electrical Engineering, Uppsala University, PO Box 65, SE-75103, Uppsala, Sweden. (email: sribalaji.anand@angstrom.uu.se}
	\thanks{A. M. H. Teixeira is with the Department of Information Technology, Uppsala University, PO Box 337, SE-75105, Uppsala, Sweden. (email: andre.teixeira@it.uu.se}
	\thanks{A. J. Gallo and R. M. G. Ferrari are with the Delft Center for Systems and Control, Mechanical, Maritime, and Materials Engineering, TU Delft, Delft, Netherlands (email:{a.j.gallo,r.ferrari}@tudelft.nl)
	}
}

\begin{document}
	
	\maketitle
	
	\begin{abstract}
		
		This paper addresses the design of an active cyber-attack detection architecture based on multiplicative watermarking,
		allowing for detection of covert attacks.
		\textcolor{ajg}{We propose an optimal design problem, relying on the so-called output-to-output $\ell_2$-gain, which characterizes the maximum gain between the residual output of a detection scheme and some performance output.
		Although optimal, this control problem is non-convex. 
		Hence, we propose an algorithm to design the watermarking filters by solving the problem suboptimally via LMIs.
		}
		We show that, against covert attacks, the output-to-output $\ell_2$-gain is unbounded without watermarking, and we provide a sufficient condition for boundedness in the presence of watermarks.


	\end{abstract}
	
	\section{Introduction}
	
	Modern engineering systems have been characterized by an ever-growing penetration of cyber resources within physical systems, embedding sensing, communication and computational capabilities due to the  reduction of costs of enabling technologies.
	The scale of the integration has lead to the study of so-called cyber-physical systems (CPS) \cite{baheti2011cyber,giraldo2017security}.
	
	Many of the systems that can be appropriately described as CPSs, such as transportation networks, electrical power grids, water distribution networks, among others, are safety critical: indeed, malfunctions in their operation may lead to lack of safety to operators or the general public, as well as economic and societal costs.
	Apart from accidental malfunctions, given the integration of cyber resources in CPS, these systems have been made the target of malicious attacks, as some high-profile cases show \cite{falliere2011w32,case2016analysis,sobczak2019dos}.
	
	This has lead to the development of secure control. 
	Differently from the research on cyber-security in information technology, secure control relies on system-theoretic approaches to protect system confidentiality, integrity and availability, and it exploits knowledge of the physical plants at the core of CPS to detect attacks.
	Control-theoretic cyber security has focused predominantly on cyber-attack detection and resilience, i.e. automatic methods to guarantee a certain level of performance against malicious interference; in this paper we focus on methods for cyber-attack detection.
	Among the methods that have been proposed in literature, a distinction may be drawn between \textit{passive} and \textit{active} detection methods, where the prior exploit measurements and knowledge of the system to detect the presence of malicious agents, while the latter actively perturb signals to improve the detectability of attacks.
	While a detailed overview of active methods is out of the scope of this paper, a few examples can be found in \cite{weerakkody2015detecting,weerakkody2019challenges,ferrari2020switching,gallo2018distributedWM}.
	Here we focus on the active detection method proposed in \cite{ferrari2020switching}, called \textit{multiplicative watermarking}.

	Presented in \cite{ferrari2020switching} and inspired by authentication schemes with weak cryptographic guarantees, multiplicative watermarking relies on a pair of linear systems, a watermark generator and remover, to modulate the information transmitted between the plant and the controller.
	This allows for detection of a number of attacks \textcolor{ajg}{without  degrading the performance of the closed-loop CPS.} 
	Indeed, the watermark generator and remover are specifically designed such that, the effect of the watermark is removed from the input and output data transmitted between the controller and the plant within the CPS.
	This improves the detection capabilities of the diagnostic tools of the CPS, as was shown in \cite{ferrari2020switching}, where the properties of  multiplicative watermarking on the output-side communication network of a CPS were investigated.
	
	In this paper we present a method for \textcolor{ajg}{(sub-)}optimal design of multiplicative watermarking units in CPS.
	Specifically, considering watermarking on both the input and the output-side communication between plant and controller, and  relying on output-to-output $\ell_2$-gain (OOG) \cite{teixeira2015strategic}, we propose an algorithm minimizing the OOG in the presence of covert attacks.
	The OOG is an index of worst case gain between the residual input of the diagnosis architecture and a performance output of the plant; as such, it is well suited to be considered as an index for optimal design of control parameters for cyber-attack detection \cite{teixeira2019optimal,anand2020joint}.
	The contributions of this paper are the following:
	\textcolor{ajg}{
		\begin{enumerate}[a.]
			\item the formulation of the design of watermarking systems based on OOG;
			\item the analysis of OOG against a covert attack \cite{smith2015covert}; 
			\item the definition of a sufficient condition determining when multiplicative watermarking improves the OOG of the closed-loop system;
			\item the design of the watermarking filters to minimize the OOG against covert attacks.
		\end{enumerate}
	}
	
	The remainder of the paper is structured as follows: in Section~\ref{ch:probFor} we present the structure of the CPS with watermarking units, define the attack strategy, and formally introduce the problem.
	Following this, in Section~\ref{ch:OOG}, we present the OOG together with some of its fundamental properties.
	Thus, in Section~\ref{ch:ctrl}, we show how the OOG may be used for optimal control design,
	\textcolor{ajg}{introducing an algorithm to compute the watermarking filters suboptimally, and some of its properties in Sections~\ref{ch:WM} and~\ref{ch:WM:prop}.}
	Finally, in Section~\ref{ch:sim} we give a numerical example. 

	\subsection*{Notation}
	Let $a:\mathbb N_+ \rightarrow \mathbb R^n$ be a real-valued discrete-time sequence. 
	Given a time horizon $[0,N] \doteq \{k \in \mathbb N_+ : 0 \leq k \leq N\}$, the $\ell_2$-norm of $a$ over $[0,N]$ is defined as: $\|a\|^2_{\ell_2,[0,N]} \doteq \sum_{k = 0}^N a[k]^\top a[k]$.
	Define $\ell_2 \doteq \{x:\mathbb N_+ \rightarrow \mathbb R^n : \|x\|_{\ell_2}^2 \doteq \|x\|_{\ell_2,[0,\infty]}^2 < \infty\}$ and the extended $\ell_2$ space $\ell_{2e} \doteq \{x:\mathbb N_+ \rightarrow \mathbb R^n : \|x\|_{\ell_2,[0,N]}^2 < \infty \,, \forall N  \in \mathbb N_+\}$.

	\section{Preliminaries and problem formulation}\label{ch:probFor}
	\subsection{System description}
\begin{figure}[t]
		\centering
		\includegraphics[width=0.9\linewidth]{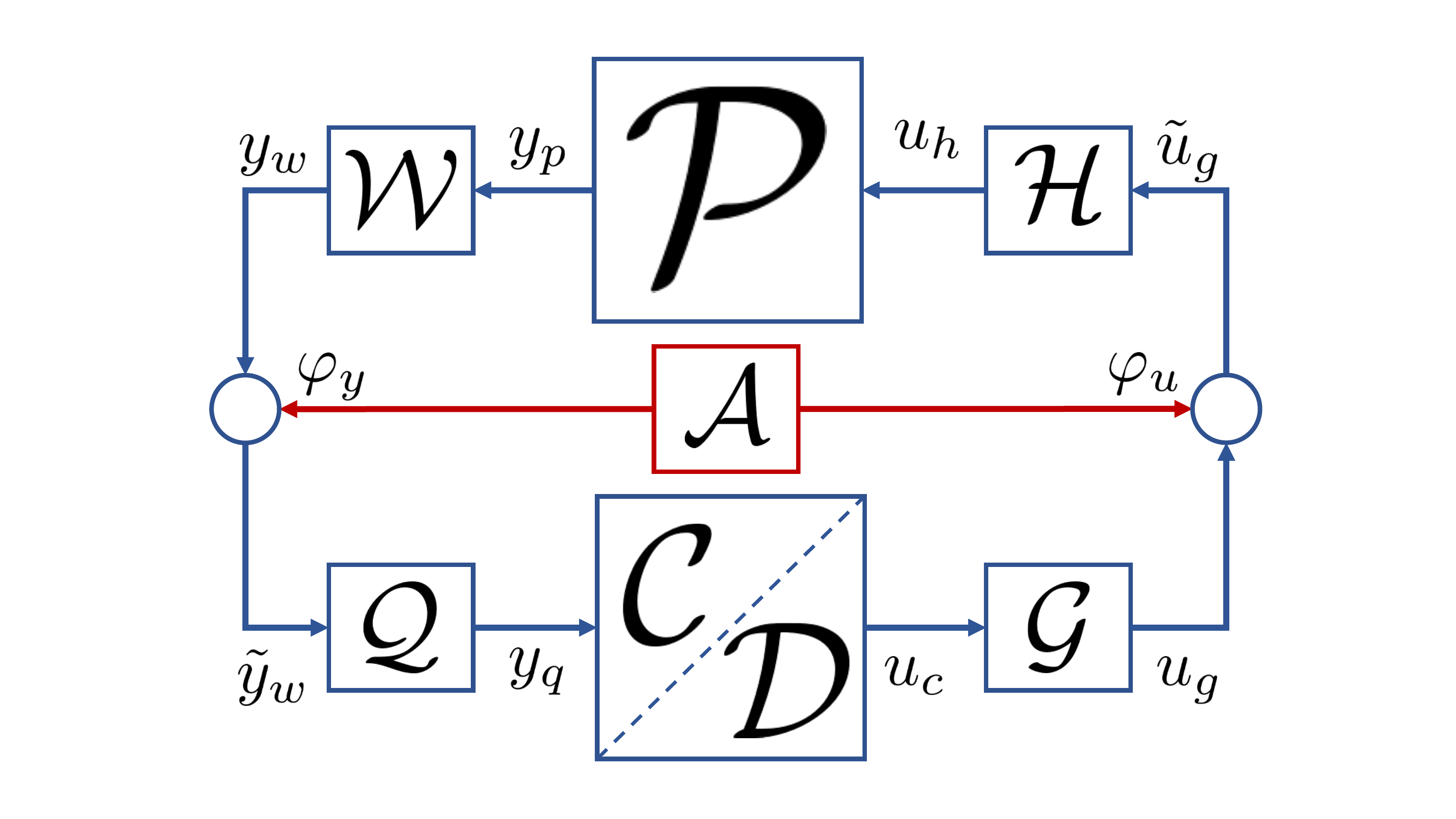}
		\caption{Block diagram of the closed-loop cyber-physical system, including plant $\PP$, controller $\CC$ and \textcolor{ajg}{watermarking filters} $\{\WW,\QQ\}$ and $\{\GG,\HH\}$.
			\textcolor{ajg}{Signals transmitted between the plant and the controller may being altered by an attacker $\mathcal A$.} 
		}
		\label{fig:sys}
	\end{figure}

	We consider a linear time-invariant (LTI) CPS as that shown in Fig.~\ref{fig:sys}, composed of a plant, $\PP$, controlled by a dynamic controller $\CC$. The control input and measurement output of the plant are transmitted between the plant and the controller over a communication network, and they are modulated through multiplicative watermarking systems \cite{ferrari2020switching}.
	
	\subsection{Plant and controller}
	We start the analysis of the closed-loop CPS by defining the physical plant and its controller.
	The physical plant $\PP$ is modeled as a discrete-time  LTI system: 
	\begin{equation}\label{eq:sys}
		\mathcal{P} : \left\{
		\begin{aligned}
			&x_p^+ = A_p x_p + B_p u_h \\
			&y_J = C_J x_p + D_J u_h\\
			&y_p = C_p x_p
		\end{aligned}
		\right.
	\end{equation}
	where $x_p \in \mathbb{R}^{n}$, $u_h \in \mathbb R^{m}$ are the plant's state and its input, which has been transmitted over the communication network and from which the watermark has been removed by $\HH$.
	The signal $y_p \in \mathbb R^{p}$ is the measured output of the system, while
	the performance of the system is evaluated over an interval $[0,N], N \in \mathbb N$, according to the cost function \cite{zhou1996robust}:
	\begin{equation}
		J(x_p,u_h) = \|C_J x_p + D_J u_h\|_{\ell_2,[0,N]}^2 = \|y_J\|_{\ell_2,[0,N]}^2
	\end{equation}
	where $y_J \in \mathbb R^{{p_J}}$ is the virtual performance output of $\PP$.
	All matrices are supposed to be of the correct dimensions.
	
	\begin{assumption}
		We assume that \textcolor{ajg}{$\mathcal P$ is} 
		such that $(A_p,B_p)$ is controllable and $(C_p,A_p)$ is  observable.
		$\hfill\triangleleft$
	\end{assumption}
	
	\textcolor{ajg}{The controller $\CC$ is defined as the following:}
	\begin{equation}\label{eq:cntrl}
		\mathcal C: 
		\begin{cases}
			\hat{x}_p^+ = A_p \hat{x}_p + B_c u_c + Ly_r\\
			u_c = K\hat{x}_p\\
			\hat{y}_p = C_p \hat{x}_p\\
			y_r = y_q - \hat{y}_p
		\end{cases}
	\end{equation}
	where $\hat{x}_p \in \mathbb{R}^{n}$
	\textcolor{ajg}{is an estimate of}
	$x_p$, 
	$u_c \in \mathbb{R}^{m}$ is the controller-\textcolor{ajg}{defined}
	input to the system, 
	$\hat{y}_p \in \mathbb{R}^{p}$ is the output estimate, and $y_r \in \mathbb{R}^{p}$ is a residual output which may be used for cyber-attack detection;
	$K$ and $L$ are chosen to optimize the closed-loop performance. 
	
	\begin{remark}
		\textcolor{ajg}{
			Definition of the controller in \eqref{eq:cntrl} assumes that the controller and the detector are colocated, and thus information available to $\CC$ may be used for detection.
		}
		$\hfill\triangleleft$
	\end{remark}
	
	We note that, in \eqref{eq:sys} and \eqref{eq:cntrl}, we have exploited control input $u_h$ and measurement $y_q$, rather than $u$ and $y_p$.
	These represent the output of the watermark remover systems $\HH$ and $\QQ$, respectively.
	As shown in the following, $\HH$ and $\QQ$ are such that in nominal conditions $u_h = u_c$ and $y_q = y_p$.

	\subsection{Watermark generation and removal}
	The watermark generators and removers are taken to be linear systems, for which series interconnection is the identity.
	Specifically, the input and output watermarking pairs are defined as \textcolor{ajg}{$\mathcal{G},\mathcal H$ and $\mathcal W, \mathcal Q$, respectively.
		Their dynamics are given by the following:
		\begin{equation}\label{eq:sys:WM}
			\Sigma: \begin{cases}
				x_\sigma^+ = A_\sigma x_\sigma + B_\sigma \nu_\sigma\\
				\gamma_\sigma = C_\sigma x_\sigma + D_\sigma \nu_\sigma
			\end{cases}
		\end{equation}
		where subscript $\sigma \in \{g,h,w,q\}$ defines whether the state, input, or output are associated with $\mathcal G, \mathcal H, \mathcal W, \mathcal Q$.
		The systems are square, and
		\begin{equation}
			\begin{split}\label{eq:WM:input}
				\matrices{\nu_g^\top, &\nu_h^\top, &\nu_w^\top, &\nu_q^\top }^\top \doteq \matrices{u_c^\top, &\tilde{u}_g^\top, &y_p^\top, &\tilde{y}_w^\top}^\top,\\
				\matrices{\gamma_g^\top, &\gamma_h^\top, &\gamma_w^\top, &\gamma_q^\top }^\top \doteq \matrices{u_g^\top, &u_h^\top, &y_w^\top, &y_q^\top}^\top,
			\end{split}
		\end{equation}
		where a tilde is added to a variable to highlight it as being transmitted over a communication network, and therefore possibly subject to attack, as per Fig.~\ref{fig:sys}.
		All matrices are of appropriate dimensions, all systems are stable.
	}
	
	
	\begin{remark}
		\textcolor{ajg}{Given that $\mathcal G, \mathcal H, \mathcal W, \mathcal Q$ are defined by the system operator, their stability can be guaranteed.}
		$\hfill\triangleleft$
	\end{remark}
	
	\textcolor{ajg}{The watermarking pairs are designed as in \cite{ferrari2020switching}, i.e.:}
	\begin{equation}
		\begin{array}{cc}
			\HH \doteq \GG^{-1} &\QQ \doteq \WW^{-1}\,,  \\
		\end{array}
	\end{equation}
	where the inverse of a system is given in Definition~\ref{def:inv:ss}.
	\begin{definition}[{\cite[Lemma 3.15]{zhou1996robust}}]\label{def:inv:ss}
		Define the transfer function from the tuple $(A,B,C,D)$ as:
		\begin{equation}
			G(z) = \left[\begin{array}{c|c}
				A &B\\
				\hline
				C &D
			\end{array}
			\right],
		\end{equation}
		and suppose that $D$ is an invertible matrix. Then
		\begin{equation}\label{eq:sys:inv}
			G^{-1}(z) = \left[\begin{array}{c|c}
				A - BD^{-1}C &-BD^{-1}\\
				\hline
				D^{-1}C &D^{-1}
			\end{array}
			\right]
		\end{equation}
		is the inverse transfer function of $G(z)$.
		$\hfill\triangleleft$
	\end{definition}
	
	Given this definition, \textcolor{ajg}{in the absence of attacks, we have:
		\begin{align}\label{eq:sys:WM:inv}
			&\begin{array}{cc}
				\mathcal{Q}(z)\mathcal{W}(z) = I_{p}\,,&  \mathcal{H}(z)\mathcal{G}(z) = I_{m}\,,
			\end{array}\\
			&\begin{array}{ccc}
				u_h[k] = u[k]\,, & y_q[k] = y_p[k]  \,, &\forall k \in \mathbb{N}_+,
			\end{array}\label{eq:sys:WM:inv:2}
		\end{align}
		assuming $x_s[0] = x_t[0], s\in\{g,w\}, t\in \{h,q\}$.
	}
	
	\subsection{Cyber-attack modeling}
	We consider a malicious agent $\mathcal A$, as in Fig.~\ref{fig:sys}, capable of injecting attacks to the signals transmitted between the controller and the plant. \textcolor{ajg}{This is formally} modeled as:
	\begin{subequations}\label{eq:atk}
		\begin{align}
			&\tilde{u}_g[k] \doteq u_g[k] + \beta_u[k-K_a^u] \varphi_u[k]\\
			&\tilde{y}_w[k] \doteq y_w[k] + \beta_y[k-K_a^y] \varphi_y[k]
		\end{align}
	\end{subequations}
	where $\varphi_u[k]$ and $\varphi_y[k]$ are actuator and sensor attack sequences defined by the malicious agent. \textcolor{ajg}{For $l \in \{u,y\}$, the function} $\beta_l[\cdot]$ is an activation function,
	defined as:
	\begin{equation}
		\begin{array}{cc}
			\beta_l [k] = \begin{cases}
				(1- b_l^k) \,, & \text{if } k \geq 0 \\
				0 \,, &\text{otherwise}
			\end{cases}\,,
		\end{array}
	\end{equation}
	and where $K_a^l > 0$ is the initial instant of attack, and $b_l \in [0,1]$; without loss of generality, we assume $b_l = 0$.
	
	We assume that $\mathcal A$ has the necessary resources (as defined in \cite{teixeira2015secure}) to leverage a \textit{covert attack} against the CPS without watermarking, i.e. that $\varphi_u$ and $\varphi_y$ satisfy:
	\begin{equation}\label{eq:atk:cov}
		\mathcal A:
		\begin{cases}
			x_a^+ = A_p x_a + B_p \varphi_u\\
			y_a = C_p x_a\\
			\varphi_y = - y_a
		\end{cases}
	\end{equation}
	with internal state $x_a \in \mathbb R^n$, and where $\varphi_u$ is arbitrarily defined by the attacker, such that $\varphi_u \in \ell_{2e}$, and with $K_a^u = K_a^y = K_a$.
	We consider that the malicious agent does not have knowledge of the watermarking systems $\{\WW,\QQ,\GG,\HH\}$.

	\subsection{Cyber-attack detection}
	
	Given the possibility of cyber-attacks, we equip the controller with detection logic.
	We use the innovation $y_r$ as a \textit{residual}, compared to an appropriately defined threshold $\theta_r$, designed to satisfy the trade-off between ability of detecting attacks and robustness against noise.
	For the purpose of this paper, the threshold is set as $\theta_r = 1$.
	Thus, for $N > 0$, the detection test may be formalized as:
	\begin{equation}
		\|y_r\|_{\ell_2,[0,N]}^2 \geq \theta_r.
	\end{equation}
	
	It is known that, in the absence of watermarking, it is sufficient for the attacker to select $x_a[K_a] = 0$ for the attack defined in \eqref{eq:atk}-\eqref{eq:atk:cov} to be stealthy, i.e. for $\varphi \doteq \matrices{\varphi_u^\top, \,\varphi_y^\top}^\top$ to not influence the residual output $y_r$ \cite{smith2015covert,barboni2020detection}.

	\subsection{Problem formulation}
	
	Having presented the overall architecture for cyber-attack detection with watermarking, we can formally present the objective of this paper.
	Let us recall the system's OOG \cite{teixeira2015strategic}:
	\textcolor{ajg}{
		introduced as a metric to quantify the effect of worst-case stealthy attacks on the performance of a system, it measures the amplification between the residual and performance outputs, $y_r$ and $y_p$, respectively.
	}
\textcolor{ajg}{	
	We introduce the following:
	\begin{equation}\label{eq:cl:generic}
		{\mathcal{S}}:
		\begin{cases}
			{x}^+ = A x + Ba\\
			y_1 = C_1 x + D_1 a\\
			y_2 = C_2 x + D_2 a
		\end{cases}
	\end{equation}
	where $x \in \mathbb R^\nu$ is the closed loop system state, $a \in \mathbb R^\mu$ is the attack signal, $y_1 \in \mathbb \mathbb R^{\upsilon_1}$ is a residual output used for detection and $y_2 \in \mathbb R^{\upsilon_2}$ is the performance output.
	The system $\mathcal S$ can be seen as any closed loop system $\PP$, $\CC$ driven by an external attack signal, with $y_r$, $y_J$ and $\varphi$ substituted with $y_1$, $y_2$ and $a$, respectively.
}
	
	\begin{definition}\label{def:o2o}
		Take $\mathcal S$ as \textcolor{ajg}{in \eqref{eq:cl:generic},}
		the output-to-output $\ell_2$-gain is defined as:
		\begin{equation}\label{eq:o2o}
			\begin{array}{lcl}
				\|\mathcal{S}\|_{\ell_2,y_2\leftarrow y_1}^2 \doteq &\displaystyle\sup_{a\in\ell_{2e}} &\|y_2\|_{\ell_2}^2 \\
				&s.t. &\|y_1\|_{\ell_2}^2 \leq 1,\quad
				x[0] = 0
			\end{array}.
		\end{equation}
		$\hfill\triangleleft$
	\end{definition}

	
	We now formulate the central problem of this paper.
	
	\begin{problem}
	Design the parameters of $\{\WW,\QQ,\GG,\HH\}$ such that \eqref{eq:sys:WM:inv} holds, while minimizing the system OOG.
	$\hfill\triangleleft$
	\end{problem}
	



	\section{Output-to-output $\ell_2$-gain}\label{ch:OOG}
	Let us briefly summarize the main results in \cite{teixeira2015strategic}, to introduce the main properties of the output-to-output $\ell_2$-gain.
	
	As shown in \cite{teixeira2015strategic}, the non-convex optimization problem \eqref{eq:o2o},
	can be cast into its convex dual
	\begin{equation}\label{eq:o2o:opt1}
		\begin{array}{lclll}
			\|{\mathcal{S}}\|_{\ell_2,y_2\leftarrow y_1}^2 \doteq &\displaystyle\inf_{\gamma > 0} &\gamma \\
			&s.t. &\|y_2\|_{\ell_2}^2 \leq \gamma \|y_1\|_{\ell_2}^2\,, \quad \textcolor{ajg}{\forall a\in \ell_{2e}}\\
			& &x[0] = 0
		\end{array}.
	\end{equation}

	Thus, recalling Definition~\ref{def:o2o}, and given that $y_1$ is taken to be a suitable residual output of $\mathcal S$, defining $\gamma^* \doteq \|{\mathcal S}\|_{\ell_2,y_2 \leftarrow y_1}^2$, note that \textcolor{ajg}{it} 
	can be interpreted as the maximum amplification of the system from $\|y_1\|_{\ell_2}^2$ to $\|y_2\|_{\ell_2}^2$, i.e.
	\begin{equation}
		\begin{array}{cc}
			\|y_2\|_{\ell_2}^2 \leq \gamma^* \|y_1\|_{\ell_2}^2 \,,&  
			x[0] = 0.
		\end{array}
	\end{equation}

	Furthermore, recalling \eqref{eq:o2o}, $\gamma^*$ also represents the worst case impact of an attack on the performance of the system:
	\begin{equation}
		\|y_2\|_{\ell_2}^2 \leq \gamma^* \|y_1\|_{\ell_2}^2 \leq \gamma^*\textcolor{ajg}{\,, \quad \forall a \in \ell_{2e}}
	\end{equation}
	Following terminology in \cite{teixeira2015strategic}, any attack capable of achieving the worst case gain $\gamma^*$, the optimal solution of \eqref{eq:o2o:opt1}, is said to be a \textit{strategic attack}.
	
	Although more readily solvable than \eqref{eq:o2o}, the optimization problem \eqref{eq:o2o:opt1} is formulated in signal space, and is therefore infinite dimensional.
	By relying on results from dissipative system theory, it can be shown\footnote{
		\textcolor{ajg}{
			We refer the interested reader to \cite{teixeira2019optimal} for details.
		}
	} 
	that the following holds.
	\begin{proposition}{\cite[Prop.1]{teixeira2019optimal}}\label{prop:OGG}
		Consider the LTI system ${\mathcal S}$ defined in \eqref{eq:cl:generic}, and presume that $(A,B)$ is controllable and $(C_1,A)$ is observable.
		Define a supply function 
		\begin{equation}\label{eq:supply}
			s(x,a) = \gamma\|y_1[k]\|_2^2 - \|y_2[k]\|_2^2
		\end{equation}
		Then the following statements are equivalent:
		\begin{enumerate}[a.]
			\item The system ${\mathcal S}$ is dissipative w.r.t. $s(x,a)$;
			\item For all trajectories of $x$, and $N > 0$ and $x[0] = 0$, \label{cond:sup}
			\begin{equation}
				\sum_{\kappa = 0}^{N-1} s(x[\kappa],a[\kappa]) \geq 0.
			\end{equation}
			\item There exists some $P \succcurlyeq 0$ such that \label{cond:opt}
			\begin{equation}
				\begin{split}
					R(P) 
					- \gamma \matrices{C_1^\top \\ D_1^\top} \matrices{C_1 \,D_1}
					+ \matrices{C_2^\top \\D_2^\top} \matrices{C_2 \,D_2}
					\preccurlyeq 0.
				\end{split}
			\end{equation}
			with $R(P)$ defined as:
			\begin{equation}\label{eq:R(P)}
				R(P) = \matrices{
					A^\top P A - P & A^\top P B \\ 
					A^\top P B     & B^\top P B
				}.
			\end{equation}
		\end{enumerate}
		$\hfill \square$
	\end{proposition}
	
	Thus, considering supply function $s(x,a)$ defined in \eqref{eq:supply}, in light of Proposition~\ref{prop:OGG}, and as shown in \cite{teixeira2015strategic}, it is possible to \textcolor{ajg}{compute} the OOG of ${\mathcal S}$ as $\gamma^* = \|{\mathcal S}\|_{\ell_2,y_2\leftarrow y_1}^2$, with 
	\begin{equation}
		\begin{array}{l}
			\begin{array}{cc}
				\gamma^* = \displaystyle \min_{P,\gamma} &\gamma  
			\end{array}\\
			\begin{array}{cll}
				s.t. &P\succcurlyeq 0 \,, \gamma > 0\\
				&R(P) - \gamma \matrices{C_1^\top \\ D_1^\top} \matrices{C_1 \,D_1}        + \matrices{C_2^\top \\D_2^\top} \matrices{C_2 \,D_2} \preccurlyeq 0
			\end{array}.
		\end{array}
	\end{equation}
	
	\begin{proposition}[{\cite[Th. 2]{teixeira2015strategic}}]\label{prop:OOG:bound}
		Consider the LTI system ${\mathcal S}$ defined in \eqref{eq:cl:generic} with OOG $\gamma^*$.
		The \textcolor{ajg}{OOG is finite} if and only if  either of the following conditions hold:
		\begin{enumerate}[a.]
			\item the system $(A,B,C_1,D_1)$ has no unstable zeros associated with reachable $x_\lambda$; \label{cond:zeros}
			\item the unstable zeros of $(A,B,C_1,D_1)$ associated with a reachable $x_\lambda$ are also zeros of $(A,B,C_2,D_2)$;
		\end{enumerate}
		where $x_\lambda \neq 0$ is the eigenvector associated to $\lambda \in \sigma(A)$.
		$\hfill\square$
	\end{proposition}
	
	
	As pointed out in \cite{teixeira2019optimal}, the OOG has a fundamental limitation, summarized in the following proposition.
	\begin{proposition}[{\cite[Lem. 1]{teixeira2019optimal}}]\label{prop:OGG:lim}
		Let $D_2 \neq 0$, full column rank, and $D_1 = 0$. Then the OOG of ${\mathcal S}$ is unbounded.
		$\hfill\square$
	\end{proposition}
	
	In light of Proposition~\ref{prop:OGG:lim}, in the remainder of this article, we assume that $D_2 = 0$, i.e. that $D_J = 0$ in \eqref{eq:sys}.
	
	\begin{remark}
		A consequence of assuming $D_2 = 0$ is that the performance cost $J(x,u)$ is evaluated as a function of the state alone.
		Although this may be restrictive in general, as it does not pose any cost on the energy required for actuation, it permits the use of OOG defined in \eqref{eq:o2o} rather than the \textit{truncated} OOG presented in \cite{teixeira2019optimal}. Analysis of optimality with the latter is left as future work.
		These limitations 
		can be remedied by augmenting the plant dynamics with actuator dynamics, or by introducing a time-delay in the application of the input, as in \cite{anand2020joint}.
		$\hfill\triangleleft$
	\end{remark}
	
	Finally, before 
	\textcolor{ajg}{designing}
	the controller matrices, it is important to note that the formulation of the OOG implicitly presumes that the attacker has full knowledge of the closed loop dynamics ${\mathcal S}$.
	As will be shown in the following, this will be fundamental when dealing with the performance of the closed-loop dynamics including watermarking.
	
	\section{Optimal controller design}\label{ch:ctrl}
	Having presented some background information on the OOG, we are now ready to discuss the design of optimal control gains $K$ and $L$ in $\CC$ under covert attacks.
	Consider the closed-loop system in \eqref{eq:cl:generic} for the system defined by the feedback interconnection of $\PP$ and $\CC$, with state $x \doteq \matrices{x_p^\top,\, e^\top}^\top$, with $e \doteq x_p - \hat{x}_p$, $a \doteq \varphi$, and $y \doteq \matrices{y_r^\top, \, y_J^\top}^\top$.
	The system matrices can be derived 
	\textcolor{ajg}{from the feedback interconnection of $\PP$ and $\CC$.}
	\textcolor{ajg}{Clearly,} the closed loop dynamics of the system depend on the definition of $K$ and $L$ in $\CC$.
	Thus, 
	they can be included as decision variables in the optimization problem defining $\|\mathcal S\|_{\ell_2,y_2\leftarrow y_1}$: 
	\begin{equation}\label{eq:ctrl:opt1}
		\begin{array}{l}
			\begin{array}{cc}
				\displaystyle \min_{P,\gamma,K,L} & \gamma
			\end{array}\\
			\begin{array}{cl}
				\text{s.t.}& P \succcurlyeq 0 \,, \, \gamma > 0\,,\\
				& \matrices{ A^\top P A - P &  A^\top P  B \\  B^\top P  A & B^\top P B} - \gamma \matrices{ C_1^\top\\  D_1^\top} \matrices{ C_1 & D_1} \\
				&+ \matrices{ C_2 & D_2}^\top \matrices{ C_2 & D_2} \preccurlyeq 0
			\end{array}
		\end{array}.
	\end{equation}
	
	\begin{theorem}\label{th:noWM:unbound}
		Consider
		$\mathcal S$ 
		subject to covert attack \eqref{eq:atk:cov}.
		The OOG is unbounded, irrespective of $K$ and $L$.
		$\hfill\square$
	\end{theorem}
	\begin{proof}
		\textcolor{ajg}{
			In the interest of space, we omit the proofs.
		}
	\end{proof}
	
	Theorem~\ref{th:noWM:unbound} shows that if the malicious agent has the capabilities of performing a covert attack, the OOG is not a suitable criterion for the design of the control matrices.
	Therefore, other approaches may be preferred for \textcolor{ajg}{its design.}

	\section{Watermarking system design}\label{ch:WM}
	
	
	Having shown that, in the absence of watermarking systems, there are no control gains $K$ and $L$ such that $\|\mathcal{S}\|_{\ell_2,y_J\leftarrow y_r}^2$ is bounded, we now show how including watermarking units \eqref{eq:sys:WM} may be used to improve the closed-loop performance against covert attacks.
	
	To this end, take the closed-loop system $\mathcal{S}$ in \eqref{eq:cl:generic} to represent the closed-loop CPS composed of the feedback interconnection of $\PP$ and $\CC$ together with the watermarking units $\{\WW,\QQ\}$ and $\{\GG,\HH\}$, as shown in Fig.~\ref{fig:sys}.
	Thus, defining $x \doteq \matrices{x_p^\top,\, e^\top,\, x_g^\top,\, x_h^\top,\, x_w^\top,\, x_q^\top,\, x_a^\top}^\top$, $a \doteq \varphi$, and $y \doteq \matrices{y_r^\top,\, y_J^\top}^\top$, the closed loop dynamics are described by \eqref{eq:cl:generic} with matrices $(A,B,C,D)$ \textcolor{ajg}{found by taking the feedback connection of the plant $\PP$, the output-side watermarking systems $\{\WW,\QQ\}$, the controller $\CC$, and the input-side watermarking pair $\{\HH,\GG\}$, as shown in Fig.~\ref{fig:sys}.}
	For ease of notation, let us define $\mathbf{W} \doteq \{\WW,\QQ,\GG,\HH\}$.
	
	In light of the discussion of the OOG in the previous section, it is possible to formulate the following optimization problem, setting the watermarking system parameters as decision variables and minimizing the OOG $\|\mathcal S\|_{\ell_2,y_2\leftarrow y_1}^2$:
	\begin{equation}\label{eq:WM:opt1}
		\begin{array}{cl}
			\displaystyle \min_{P,\gamma,\mathbf{W}} & \gamma\\
			\text{s.t.} & P \succcurlyeq 0\,,\, \gamma > 0\\
			& R(P) - \gamma\matrices{C_1,\,D_1}^\top\matrices{C_1,\,D_1} \\
			&\hspace{.75cm}+ \matrices{C_2,\,D_2}^\top\matrices{C_2,\,D_2} \preccurlyeq 0.
		\end{array}
	\end{equation}
	with $R(P)$ defined in \eqref{eq:R(P)}, and $C_1$, $C_2$, $D_1$, and $D_2$ are such that $C = \matrices{C_1^\top,\, C_2^\top}^\top$ and $D = \matrices{D_1^\top,\, D_2^\top}^\top$.
\textcolor{ajg}{
    Because of the definition of the closed loop matrices $(A,B,C,D)$ with respect to the watermarking parameters, as well as the definition of $R(P)$, problem \eqref{eq:WM:opt1} is non-convex.
	Thus, to solve it \textit{suboptimally} via an LMI formulation, we consider an alternating minimization algorithm, in which the solution is found iteratively by solving for $P$ while fixing the other decision variables, and then solving for the parameters of $\mathcal S$ with the value of $P$ fixed, until a stopping criterion is met \cite{li2019alternating}.
	This leads to LMI constraints when solving for $P$, although not when solving for the parameters of $\mathcal S$.
}

	To avoid this, we consider that the watermark \textcolor{ajg}{systems}
	have some predefined \textit{structure}, 
	i.e. that some of the matrices defining them are decided \textit{a priori}, to linearize the constraints of \eqref{eq:WM:opt1} when $P$ is fixed.
	A number of different approaches may be taken, such as defining finite impulse or infinite impulse response (FIR and IIR) filters for each of the components of $u_c$ and $y_m$, or defining $D_s$ and 
	$C_s$ \textit{a priori} as $\bar{D}_s$ and $\bar{C}_s$, respectively, for $s \in \{h,q\}$.
	
	Thus, the optimization problem \eqref{eq:WM:opt1} is redefined, taking the Schur complement of $R(P)$:
	\begin{subequations}\label{eq:WM:opt:OOG}
		\begin{align}
			\displaystyle \min_{P,\gamma,\mathbf{W},P_q,P_h } & \gamma\\
			\text{s.t.} \hspace{.55cm} & P \succcurlyeq 0\,,\, \gamma > 0\,,\, P_q \succcurlyeq 0 \,,\, P_h \succcurlyeq 0\\
			&\matrices{-P_s &A_s^\top P_s \\ P_s A_s & -P_s} \preccurlyeq 0 \quad s \in \{h,q\} \label{eq:WM:opt:OOG:stab}\\
			& A_w = A_q - B_qD_q^{-1}C_q\,, \label{eq:WM:opt:OOG:inv1}\\
			& A_g = A_h - B_hD_h^{-1}C_h\,, \label{eq:WM:opt:OOG:inv2}\\
			& \begin{array}{r}
				\matrices{
					-P    & 0   & A^\top P \\ 
					0     & 0   & B^\top P \\
					P A   & P B & -P       \\
				} - \gamma \matrices{ C_1^\top\\  D_1^\top \\ 0} \matrices{ C_1 &  D_1 & 0} \\
				+ \matrices{ C_2^\top\\  D_2^\top \\ 0} \matrices{ C_2 &  D_2 & 0} \preccurlyeq 0,
			\end{array}\\
			&C_s = \bar{C}_s,\quad D_s = \bar{D}_s,\quad s \in \{h,q\},
		\end{align}
	\end{subequations}
	\textcolor{ajg}{
		where \eqref{eq:WM:opt:OOG:stab} is included, together with positive semidefiniteness of $P_s, s \in \{h,q\}$, to guarantee the stability of the watermarking systems, and \eqref{eq:WM:opt:OOG:inv1}--\eqref{eq:WM:opt:OOG:inv2} guarantee that \eqref{eq:sys:WM:inv} hold, and therefore that $\{\GG,\HH\}$ and $\{\WW,\QQ\}$ are indeed watermarking pairs.
	}
	Notice that \eqref{eq:WM:opt:OOG} is bilinear in the constraints. {Algorithm~\ref{alg:WM:OOG}} solves the problem suboptimally, by implementing an alternating algorithm \cite{li2019alternating}.

	
	\begin{algorithm}[t]
		\caption{Watermark  design}
		\begin{algorithmic}[1]
			\State \textbf{Input:} Stabilizing $K$, $L$, stable $\mathbf{W}$, $\epsilon \in \mathbb R_+$
			\State \textbf{Output:} $\mathbf{W}^*, \gamma^*, P^*$
			\vspace{.1cm}
			\hrule
			\vspace{.1cm}
			\State Set $k = 0$, $\gamma_0 = 0$, $\gamma_{-1} = \infty$, $\mathbf{W}_0 = \{\WW,\QQ,\HH,\GG\}$
			\While {$\|\gamma_k - \gamma_{k-1}\| > \epsilon$}
			\State Find $P_{k+1}$ optimizing \eqref{eq:WM:opt:OOG} w.r.t. $P$ with $\mathbf{W} = \mathbf{W}_k$;
			\State Find $\mathbf{W}_{k+1},,\gamma_{k+1}$ optimizing \eqref{eq:WM:opt:OOG} w.r.t. $\gamma$, $\mathbf{W}$, $P_q$, $P_h$ with $P = P_{k+1}$;
			\State $k = k+1$;
			\EndWhile
			\State \textbf{return:} $\mathbf{W}^*=\mathbf{W}_k,\gamma^*=\gamma_k,P^*=P_k$.
		\end{algorithmic}
		\label{alg:WM:OOG}
	\end{algorithm}

	\begin{remark}
	    \textcolor{ajg}{Note that, to achieve the worst case gain $\gamma^*$, the attacker must have full knowledge of the watermarking systems' parameters.
	    }
		$\hfill\triangleleft$
	\end{remark}

	\section{Structural constraints on solvability of Algorithm~\ref{alg:WM:OOG}}\label{ch:WM:prop}
	
	Let us briefly comment on the feasibility of {Algorithm~\ref{alg:WM:OOG}}, given the covert attack scenario considered in \eqref{eq:atk} with covert attack defined in \eqref{eq:atk:cov}. 
	If the algorithm is infeasible, this implies that there exists an input sequence $\varphi_u \in \ell_{2e}$ such that the covert attack strategy defined in \eqref{eq:atk:cov} remains undetectable in the presence of the  watermarking units.
	
	Given the structure of $\CC$, an attack that is undetectable will also guarantee $y_q^a = 0$, where $y_q^a$ is the component of $y_q$ driven by the attack $\varphi_u$.
	In turn, given invertibility of $\QQ$, it is possible to see that, if $\varphi_u$ is an undetectable sequence, then $\tilde{y}_w^a \doteq y_w^a - y_a = 0$ for all $k \geq K_a$, where, again, superscript $a$ is used to define the component of $y_w$ driven by the attack input $\varphi_u$.
	Formally, $\tilde y_w^a$ can be seen as the output of $\mathcal S^a$:
	\begin{equation}\label{eq:cl:atk}
		\mathcal S^a:
		\begin{cases}
			x^{a+} = A^a x^a + B^a \varphi_u\\
			y^a = C^a x^a + D^a \varphi_u
		\end{cases}
	\end{equation}
	where $x^a \doteq \matrices{x_h^\top,\, x_p^\top,\, x_w^\top,\, x_a^\top}^\top$, $y^a \doteq \tilde{y}_w^a$, and with matrices defined
	\textcolor{ajg}{considering the series connection of $\HH,\PP,\WW$ and the attacker-defined system $\mathcal A$ \eqref{eq:atk:cov}.}
	
	It is well known that an attack $\varphi_u \neq 0$ against $\mathcal S^a$ is undetectable, and therefore is such that $y^a = 0$, if and only if it is a zero-dynamics attack, i.e. it satisfies:
	\begin{equation}\label{eq:cl:atk:Rosn}
		\matrices{\lambda I - A^a & -B^a \\ C^a &D^a }\matrices{\bar{x}^a\\\bar{\varphi}_u} = \matrices{0\\0}
	\end{equation}
	for some $\lambda \in \mathbb C$ and some $\bar x^a$ and $\bar\varphi_u$ \cite{pasqualetti2013attack}.
	$G_w(z), G_p(z), G_h(z), G_a(z)$ are the transfer functions of $\WW,\PP,\HH$ and $\mathcal A$, respectively, with $G_a(z) \equiv G_p(z), \forall z \in \mathbb C$.
	\begin{lemma}\label{lem:WM:zeroDYn}
		Consider \eqref{eq:cl:atk}. 
		The following are equivalent:
		\begin{enumerate}[a.]
			\item Exists $\lambda \in \mathbb{C}$, $\bar{x}^a\in \mathbb{R}^n$, $\bar{\varphi}_u \in \mathbb R^m$: \eqref{eq:cl:atk:Rosn} holds;
			\item Exists $\mu \in \mathbb C$: $G_w(\mu)G_p(\mu)G_h(\mu) \equiv G_p(\mu)$, $\varphi \neq 0$.
			$\hfill\square$
		\end{enumerate}
	\end{lemma}
	\begin{proof}
		\textcolor{ajg}{
			In the interest of space, we omit the proofs.
		}
	\end{proof}
	
	
	\textcolor{ajg}{
		We now present the main theoretical result of this paper: a sufficient condition under which Algorithm~\ref{alg:WM:OOG} is feasible, guaranteeing the output-to-output gain of the closed-loop system with watermarking, $\gamma$, is finite.
	}
	

	\begin{theorem}
		Consider 
		$\mathcal S$ in \eqref{eq:cl:generic}. 
		Suppose that $\PP$ defined in \eqref{eq:sys} satisfies condition~\ref{cond:zeros}. in Proposition~\ref{prop:OOG:bound}.
		Thus, {Algorithm~\ref{alg:WM:OOG}} returns a solution such that $\gamma<\infty$, so long as the initial choice of $\mathbf{W}$ is such that \eqref{eq:WM:opt:OOG} is feasible.
		$\hfill\square$
	\end{theorem}
	\begin{proof}
	    \textcolor{ajg}{
			In the interest of space, we omit the proofs.
		}
	\end{proof}

	\section{Numerical example}\label{ch:sim}
	
	In this section, the effectiveness of the proposed {Algorithm \ref{alg:WM:OOG}} is depicted through a numerical example. Consider a plant $\PP$ and controller $\CC$ with the following parameters: $A_p=\begin{bmatrix}
    0.9191  &  0.3277\\
   -0.0768  &  0.4269
\end{bmatrix}, B_p =\begin{bmatrix}
0 \\ 1
\end{bmatrix}, C_J = \begin{bmatrix}
2 \\ 0
\end{bmatrix}^\top, {D_J =0}, C_p = \begin{bmatrix}
1 \\ 0
\end{bmatrix}^\top, K= \begin{bmatrix}
-0.3405 \\ -0.3987
\end{bmatrix}^\top,$ and $L= \begin{bmatrix}
0.5956 \\ -0.0253
\end{bmatrix}$. We consider a watermark remover at the output and the input with the structure $B_f = C_f = D_f =1, f \in \{q,h\}$.
Note that here we consider a more constrained case than the general case, where either $B_f$ or $C_f$ must be known \textit{a priori}.
The watermark generator and remover are related by \eqref{eq:WM:opt:OOG:inv1}--\eqref{eq:WM:opt:OOG:inv2}. 
We initialize the algorithm with $\epsilon = 10^{-5}$ and the stable watermark state-transition matrices $A_q=0.6714$ and $A_h=0.5201$. Firstly, the plant $\PP$ has no unstable zeros and hence the limitation discussed in Section \ref{ch:WM:prop} does not apply for the system in consideration. That is, there does not exist an input sequence which is $0$-stealthy to the detector. Although, when the watermarking scheme is absent, \eqref{eq:WM:opt:OOG} will be unbounded since the adversary knows the system. 

\begin{figure*}[ht]
	\centering
	\begin{minipage}{.32\textwidth}
		\centering
		\includegraphics[width=\linewidth]{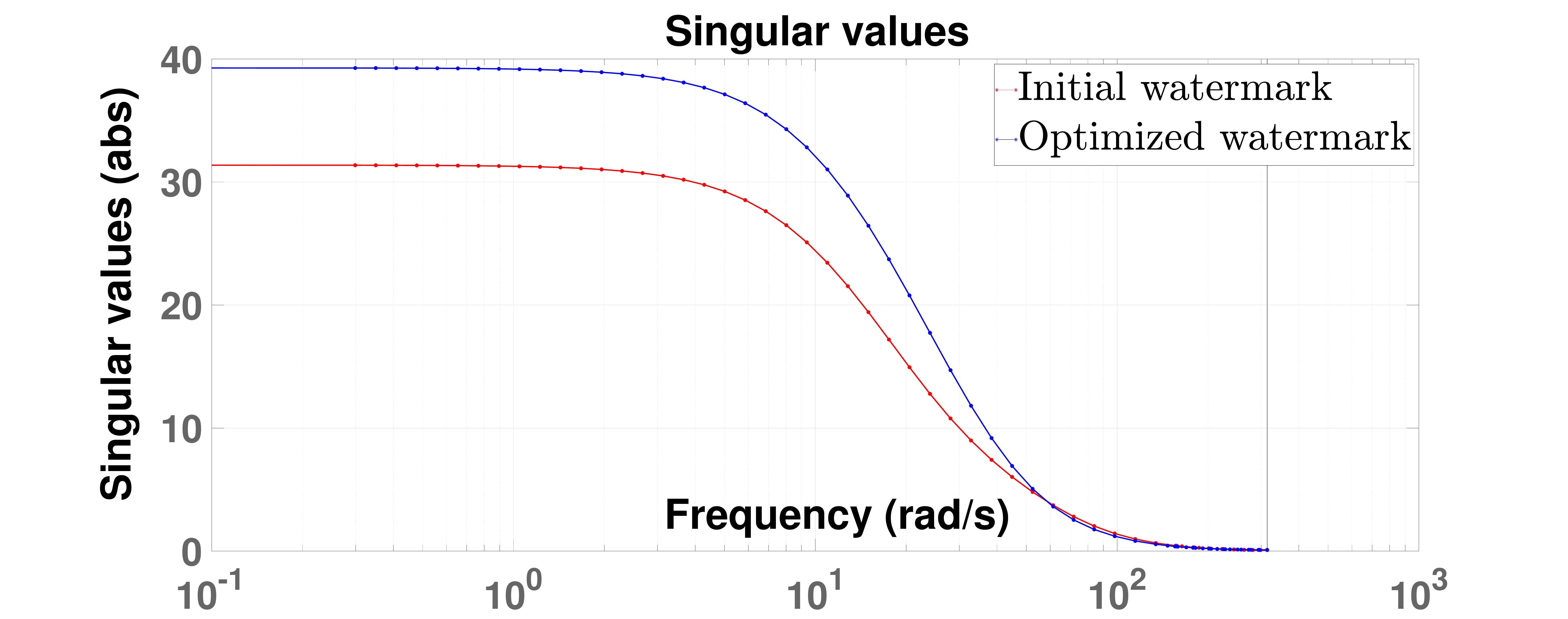}
		\caption{Singular values for the performance output}
		\label{fig:energy:perf}
	\end{minipage}
	\begin{minipage}{.32 \textwidth}
		\centering
		\includegraphics[width=\linewidth]{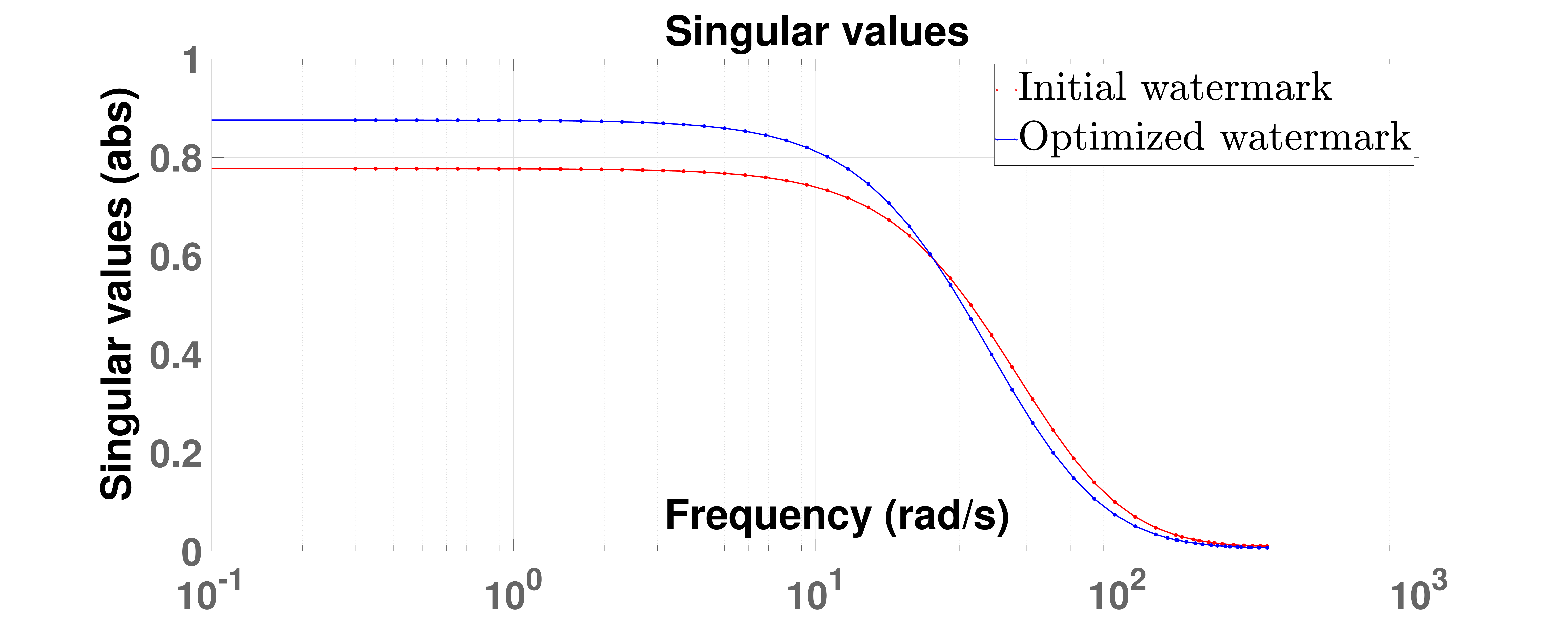}
		\caption{Singular values for the detection output }
		\label{fig:energy:de}
	\end{minipage}
	\begin{minipage}{.32 \textwidth}
		\centering
		\includegraphics[width=\linewidth]{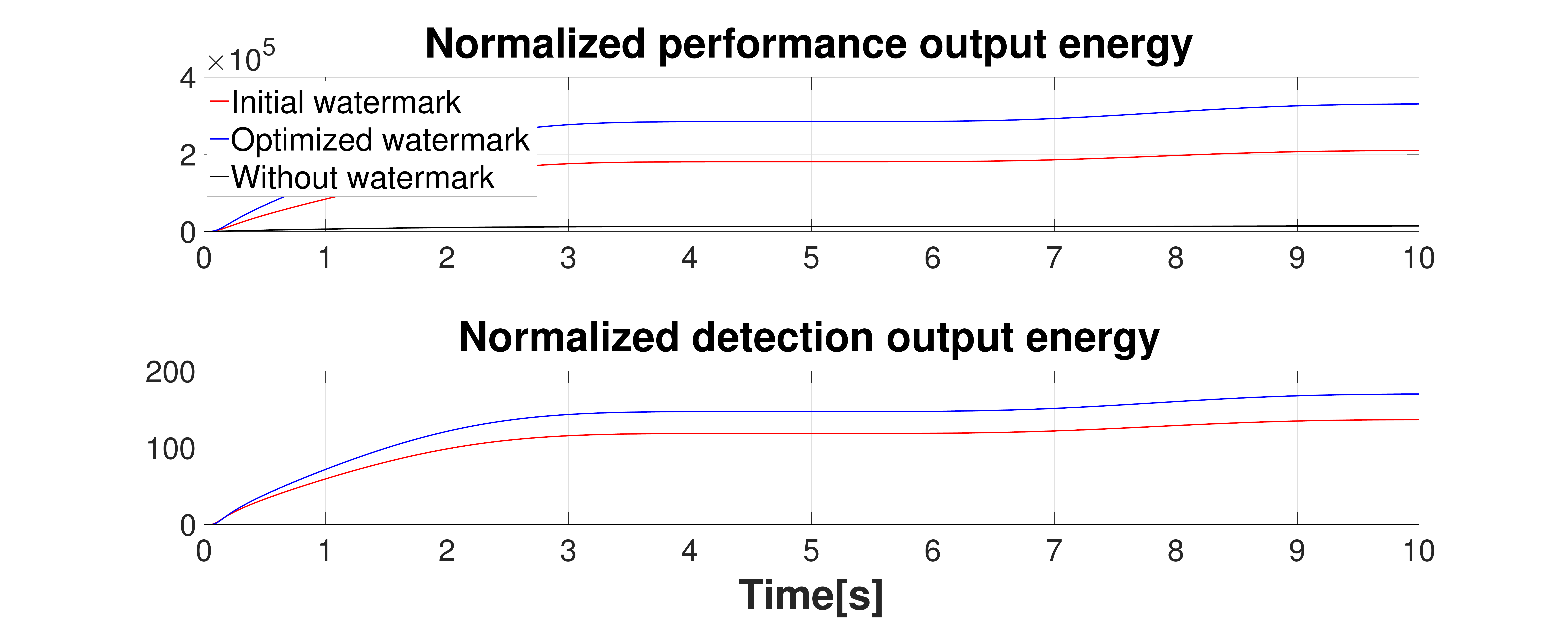}
		\caption{System outputs for a covert attack}
		\label{fig:energy}
	\end{minipage}
\end{figure*}

The objective of the OOG is to design the watermark adder and remover such that the gain (or singular values (SVs)) of the system from the attack input to the performance output is decreased, whilst the gain of the system from the attack input to the detection output is increased at all frequencies. To this end we represent the SVs, on the unit circle of the complex plane, of the system from the attack input to performance and detection outputs in Fig.~\ref{fig:energy:perf} and  Fig.~\ref{fig:energy:de}. It is evident from Fig.~\ref{fig:energy:perf} that, from $\omega = 0 \;\mbox{rad/s}$ to $\omega \approx 25 \;\mbox{rad/s}$, the performance of the system deteriorates as the SVs increase on optimizing. But the algorithm compensates for this deterioration by simultaneously increasing the detection performance as can be seen in Fig.~\ref{fig:energy:de}. The SVs of both the systems do not change much from $\omega \approx 25 \;\mbox{rad/s}$. This is because, the output-to-output gain focuses on improving the detection performance only when the performance loss is significant and vice versa. This was also pointed out in \cite{anand2020joint}.

Let us now consider an attack signal of the form \eqref{eq:atk:cov} where $\varphi_u \doteq 5 + 5 sin(k)$. The normalized energies of the performance and detection outputs with the initial watermark parameters and the optimized watermark parameters (obtained from {Algorithm \ref{alg:WM:OOG}}) are shown in Fig.~\ref{fig:energy}. The attack is undetectable in the absence of watermarks. 
Furthermore, while the effect of the attack on the performance output is increased, the effect of the attack on the detection output is increased by at least $25\%$.





	\section{Conclusions}
	In this work we have presented the optimal design of multiplicative watermarking based on the OOG of systems.
	We show how, by including multiplicative watermarking on the input and output channels of the system, its OOG can be made finite in the presence of covert attacks.
	As future work, we wish to further study structural conditions under which multiplicative watermarking may bound the OOG of the system in the presence of covert attacks.
	Further analysis is also required to \textcolor{ajg}{solve the design procedure optimally.}


	\bibliographystyle{IEEEtran}
	\bibliography{CDC21_resub_V2}

\begin{thebibliography}{10}
\providecommand{\url}[1]{#1}
\csname url@samestyle\endcsname
\providecommand{\newblock}{\relax}
\providecommand{\bibinfo}[2]{#2}
\providecommand{\BIBentrySTDinterwordspacing}{\spaceskip=0pt\relax}
\providecommand{\BIBentryALTinterwordstretchfactor}{4}
\providecommand{\BIBentryALTinterwordspacing}{\spaceskip=\fontdimen2\font plus
\BIBentryALTinterwordstretchfactor\fontdimen3\font minus
  \fontdimen4\font\relax}
\providecommand{\BIBforeignlanguage}[2]{{%
\expandafter\ifx\csname l@#1\endcsname\relax
\typeout{** WARNING: IEEEtran.bst: No hyphenation pattern has been}%
\typeout{** loaded for the language `#1'. Using the pattern for}%
\typeout{** the default language instead.}%
\else
\language=\csname l@#1\endcsname
\fi
#2}}
\providecommand{\BIBdecl}{\relax}
\BIBdecl

\bibitem{baheti2011cyber}
R.~Baheti and H.~Gill, ``Cyber-physical systems,'' \emph{The impact of control
  technology}, vol.~12, pp. 161--166, 2011.

\bibitem{giraldo2017security}
J.~Giraldo, E.~Sarkar, A.~A. Cardenas, M.~Maniatakos, and M.~Kantarcioglu,
  ``Security and privacy in cyber-physical systems: A survey of surveys,''
  \emph{IEEE Design \& Test}, vol.~34, no.~4, pp. 7--17, 2017.

\bibitem{falliere2011w32}
N.~Falliere, L.~O. Murchu, and E.~Chien, ``W32. stuxnet dossier,'' \emph{White
  paper, Symantec Corp., Security Response}, vol.~5, no.~6, p.~29, 2011.

\bibitem{case2016analysis}
R.~M. Lee, M.~J. Assante, and T.~Conway, ``Analysis of the cyber attack on the
  {U}krainian power grid,'' \emph{Electricity Information Sharing and Analysis
  Center (E-ISAC)}, 2016.

\bibitem{sobczak2019dos}
B.~Sobczak, \emph{Denial of Service attack caused grid cyber disruption:
  {DOE}}.\hskip 1em plus 0.5em minus 0.4em\relax Environment {\&} Energy
  Publishing, 2019.

\bibitem{weerakkody2015detecting}
S.~Weerakkody and B.~Sinopoli, ``Detecting integrity attacks on control systems
  using a moving target approach,'' in \emph{2015 54th IEEE Conf. on Decision
  and Contr. (CDC)}.\hskip 1em plus 0.5em minus 0.4em\relax IEEE, 2015, pp.
  5820--5826.

\bibitem{weerakkody2019challenges}
------, ``Challenges and opportunities: Cyber-physical security in the smart
  grid,'' in \emph{Smart Grid Control}.\hskip 1em plus 0.5em minus 0.4em\relax
  Springer, 2019, pp. 257--273.

\bibitem{ferrari2020switching}
R.~M. Ferrari and A.~M. Teixeira, ``A switching multiplicative watermarking
  scheme for detection of stealthy cyber-attacks,'' \emph{IEEE Trans. on
  Automat. Contr.}, 2020.

\bibitem{gallo2018distributedWM}
A.~J. Gallo, M.~S. Turan, F.~Boem, G.~Ferrari-Trecate, and T.~Parisini,
  ``Distributed watermarking for secure control of microgrids under replay
  attacks,'' \emph{IFAC-PapersOnLine}, vol.~51, no.~23, pp. 182--187, 2018.

\bibitem{teixeira2015strategic}
A.~Teixeira, H.~Sandberg, and K.~H. Johansson, ``Strategic stealthy attacks:
  the output-to-output $\ell_2$-gain,'' in \emph{2015 54th IEEE Conf. on
  Decision and Contr. (CDC)}.\hskip 1em plus 0.5em minus 0.4em\relax IEEE,
  2015, pp. 2582--2587.

\bibitem{teixeira2019optimal}
A.~M. Teixeira, ``Optimal stealthy attacks on actuators for strictly proper
  systems,'' in \emph{2019 IEEE 58th Conf. on Decision and Contr. (CDC)}.\hskip
  1em plus 0.5em minus 0.4em\relax IEEE, 2019, pp. 4385--4390.

\bibitem{anand2020joint}
S.~C. Anand and A.~Teixeira, ``Joint controller and detector design against
  data injection attacks on actuators,'' in \emph{IFAC world congress 2020
  (Virtual), Berlin, Germany, July 11-17}, 2020.

\bibitem{smith2015covert}
R.~S. Smith, ``Covert misappropriation of networked control systems: Presenting
  a feedback structure,'' \emph{IEEE Contr. Systems}, vol.~35, no.~1, pp.
  82--92, 2015.

\bibitem{zhou1996robust}
K.~Zhou, J.~C. Doyle, K.~Glover \emph{et~al.}, \emph{Robust and optimal
  control}.\hskip 1em plus 0.5em minus 0.4em\relax Prentice hall New Jersey,
  1996, vol.~40.

\bibitem{teixeira2015secure}
A.~Teixeira, I.~Shames, H.~Sandberg, and K.~H. Johansson, ``A secure control
  framework for resource-limited adversaries,'' \emph{Automatica}, vol.~51, pp.
  135--148, 2015.

\bibitem{barboni2020detection}
A.~Barboni, H.~Rezaee, F.~Boem, and T.~Parisini, ``Detection of covert
  cyber-attacks in interconnected systems: a distributed model-based
  approach,'' \emph{IEEE Trans. on Automat. Contr.}, vol.~65, no.~9, pp.
  3728--3741, 2020.

\bibitem{li2019alternating}
Q.~Li, Z.~Zhu, and G.~Tang, ``Alternating minimizations converge to
  second-order optimal solutions,'' in \emph{International Conf. on Machine
  Learning}.\hskip 1em plus 0.5em minus 0.4em\relax PMLR, 2019, pp. 3935--3943.

\bibitem{pasqualetti2013attack}
F.~Pasqualetti, F.~D{\"o}rfler, and F.~Bullo, ``Attack detection and
  identification in cyber-physical systems,'' \emph{IEEE Trans. on Automat.
  Contr.}, vol.~58, no.~11, pp. 2715--2729, 2013.

\end{thebibliography}
	
\end{document}